\documentclass[12pt]{iopart}
\usepackage[dvips]{graphicx}
\usepackage{cite}
\usepackage{iopams}
\usepackage{subfigure}
\usepackage{wasysym}

\newcommand{\be}{\begin{equation}}
\newcommand{\ee}{\end{equation}}
\newcommand{\bea}{\begin{eqnarray}}
\newcommand{\eea}{\end{eqnarray}}

\newcommand{\nn}{\nonumber}

\newcommand{\ket}[1]{|#1\rangle} 
\newcommand{\bra}[1]{\langle#1|} 
\newcommand{\eka}{\left[}
\newcommand{\ekz}{\right]}
\newcommand{\rka}{\left(}
\newcommand{\rkz}{\right)}
\newcommand{\gka}{\left\lbrace}
\newcommand{\gkz}{\right\rbrace}
\usepackage{bbold}
\newcommand{\eye}{\mathbb{1}}

\begin{document}

\title{Bloch-Zener oscillations}

\author{B. M. Breid, D. Witthaut, and H. J. Korsch}

\address{FB Physik, Technische Universit\"at Kaiserslautern,
D--67653 Kaiserslautern, Germany}

\ead{korsch@physik.uni-kl.de}

\begin{abstract}

\noindent
It is well known that a particle in a periodic potential 
with an additional constant force 
performs Bloch oscillations.
Modulating every second period of the potential, 
the original Bloch band splits into two subbands. 
The dynamics of quantum particles shows a coherent superposition of Bloch oscillations and Zener tunneling between the subbands, 
a Bloch-Zener oscillation. 
Such a system is modelled by a tight-binding Hamiltonian, a system of two minibands with an easily controllable gap. 
The dynamics of the system is investigated by using an algebraic ansatz leading to a differential equation of Whittaker-Hill type. It is shown that the parameters of the system can be tuned to generate a periodic reconstruction of the wave packet and thus of the occupation probability. As an application, the construction of a matter wave beam splitter and a Mach-Zehnder interferometer is briefly discussed.
\end{abstract}


\maketitle

\section{Introduction}

Since the beginning of quantum mechanics, the dynamics of a particle
in a periodic potential is a subject of fundamental interest.
The eigenenergies of such a system form the famous Bloch bands, where
the eigenstates are delocalized states and lead to transport \cite{Bloc28}.
If an additional static field $F$ is introduced, the eigenstates become localized.
Counterintuitively, transport is dramatically reduced and an oscillatory
motion is found, the famous Bloch oscillation.
During the last decade these Bloch oscillations have been experimentally
observed for various systems: electrons in semiconductor superlattices \cite{Feld92},
cold atoms in optical lattices \cite{Daha96}, light pulses in photonic crystals
\cite{Pert99,Mora99a} and even mechanical systems may Bloch-oscillate \cite{Guti05}.

If the external field is strong enough, tunneling between Bloch bands
becomes possible, as already discussed by Zener long ago \cite{Zene34}.
In general, successive Zener tunneling to even higher bands will lead to
decay as the bandgaps usually decrease with increasing energy.
This decay can be observed as pulsed output for a Bose-Einstein condensate in an optical lattice \cite{ande98}, for electrons in semiconductor superlattices \cite{03lifetime} or for light in waveguide arrays \cite{Trom06a}.
In this case a description in terms of Wannier--Stark resonances is more
suited than the common Bloch band picture \cite{02wsrep}.

In order to study and exploit the interplay between Bloch oscillations
and Zener tunneling, a system of at least two bands is needed, which are 
well separated from all other bands. This is achieved by introducing
another weak potential with a doubled period length.
Due to this pertubation, the ground band splits into two minibands
and Zener tunneling between the minibands must be taken into account.

In this paper we discuss the dynamics in such a period-doubled potential 
within the tight-binding approximation. We assume that the ground state
energies of the potential wells are alternately increased
or decreased by the additional potential. The Hamiltonian then reads

\be
\fl
\hspace{3mm}
\hat H=-\frac{\Delta}{4}\sum\limits_{n=-\infty}\limits^{\infty}(|n\rangle \langle n+1|+|n \rangle \langle n-1|) 
+\frac{\delta}{2}\sum\limits_{n=-\infty}\limits^{\infty}(-1)^n |n\rangle \langle n |
+Fd\sum\limits_{n=-\infty}\limits^{\infty} n |n\rangle \langle n | \,, \label{S}
\ee
where $|n\rangle$ is the Wannier state localized in the $n$th potential well.
Furthermore, $\Delta$ denotes the width of the unperturbed band ($\delta=0$), $\delta$ gives
the gap between the two minibands at the edge of the Brillouin zone and $F$ is the external force.
For $\delta=0$ this Hamiltonian reduces to the celebrated single-band
tight-binding model, which has been studied in great detail (see \cite{04bloch1d}
for a recent review).
The period-doubled tight-binding model was introduced for the field-free
case $F=0$ already in 1988 \cite{Kova88}. First pertubation theoretical
results with an external field can be found in \cite{Zhao91a,Zhao97}.
Numerical results for an analogous two-miniband system in two spatial dimensions
were reported in \cite{03bloch2D,04bloch2d}.

This paper is organized as follows: 
First of all we repeat some results for the field free system $F=0$.
The spectral properties for the case $F\neq0$ are discussed in section
\ref{sec-spectrum}, and the dynamics is analyzed in detail in section
\ref{sec-dynamics}.
We present an algebraic ansatz for the time evolution operator and analyze
the time evolution by general arguments. The interesting case of periodic, i.e. reconstructing, Bloch-Zener
oscillations and thus reconstructing occupation probabilities is analyzed in detail. These
theoretical considerations are illustrated by some numerical results.
As an application of Bloch-Zener oscillations, a beam splitting mechanism for matter waves is described and
simulated numerically. The possibility of constructing a Mach-Zehnder interferometer based on Bloch-Zener
oscillations is briefly discussed.

\section{Bloch bands in the field-free case}

\begin{figure}[htb]
\centering
\includegraphics[height=6.5cm , angle=0]{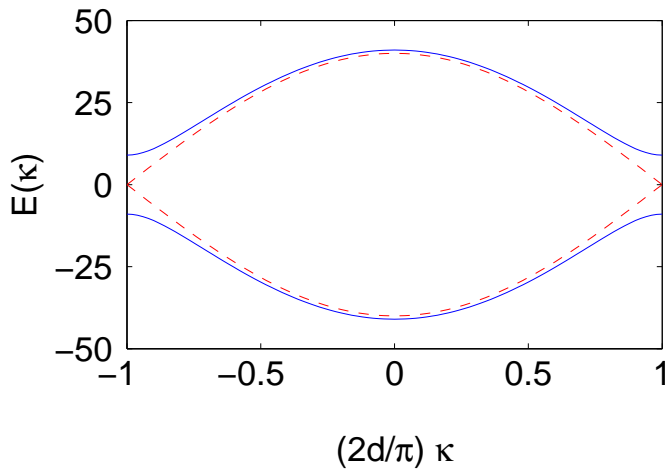}
\caption{\label{2_1} 
Dispersion relation of the double-periodic system in the reduced Brillouin zone for
$\Delta=80$. The Bloch bands for $\delta=0$ (dashed curve) split into two minibands with gap $\delta$ (solid lines for $\delta=18$).}
\end{figure}

First of all we discuss the spectrum of the period-doubled Hamiltonian (\ref{S})
for the field-free case $F=0$, i.e. Bloch bands and Bloch waves.
A straightforward calculation yields the dispersion relation
\be
  E_{\beta,\kappa}=\frac{\gamma}{2}(-1)^{\beta+1}\sqrt{\delta^2+\Delta^2\cos^2(\kappa d)},
  \label{G2_14}
\ee
with the miniband index $\beta=0,1$ and $\gamma={\rm sgn}(\delta)$,
which is illustrated in figure \ref{2_1}. For $\delta \ne 0$, the Bloch band
splits into two minibands with band gap $\delta$.

\begin{figure}[htb]
\centering
\includegraphics[height=6.5cm, width=13cm ,  angle=0]{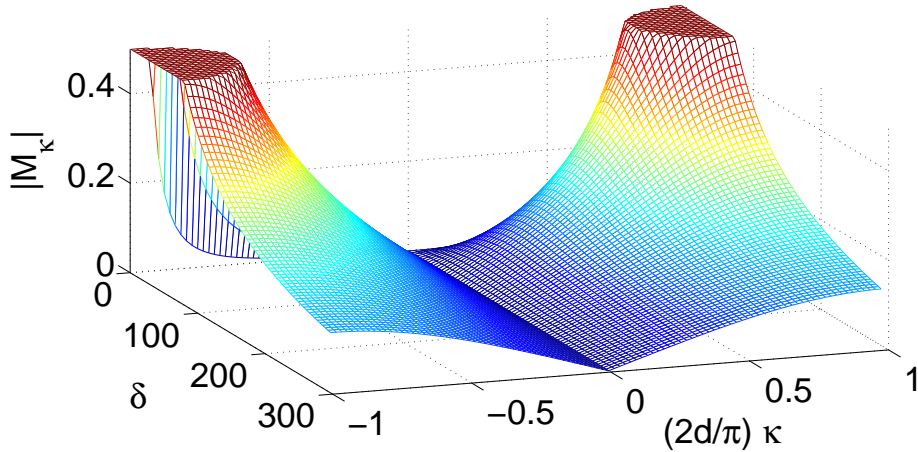}
\caption{\label{2_2_a}
Absolute value of the reduced transition matrix element
$|M_\kappa|$ for $\Delta=80$ and $dF=1$. The range $|M_\kappa|>0.5$
has been cut off to increase the visibility.}
\end{figure}

\noindent The Bloch waves $\ket{\chi_{\beta,\kappa}}$ of both bands are given by
\bea
\ket{\chi_{0,\kappa}}&=\frac{1}{\sqrt{N_\kappa}}\sum\limits_{n=-\infty}\limits^{\infty} u_\kappa \,\rme^{\rmi (2n+1) \kappa d} \ket{2n}
             +\frac{1}{\sqrt{N_\kappa}}\sum\limits_{n=-\infty}\limits^{\infty} v_\kappa \,\rme^{\rmi (2n+2) \kappa d} \ket{2n+1} \label{G2_18} \\
\ket{\chi_{1,\kappa}}&=\frac{1}{\sqrt{N_\kappa}}\sum\limits_{n=-\infty}\limits^{\infty} v_\kappa \,\rme^{\rmi (2n)   \kappa d} \ket{2n}
             -\frac{1}{\sqrt{N_\kappa}}\sum\limits_{n=-\infty}\limits^{\infty} u_\kappa \,\rme^{\rmi (2n+1) \kappa d} \ket{2n+1} \label{G2_19}
\eea
with normalization constant $N_\kappa = \pi (u_\kappa^2+v_\kappa^2)/d$ and coefficents
\be
  u_{\kappa}=\Delta \cos(\kappa d)  \quad \mbox{and} \quad
  v_{\kappa}=\delta + \gamma\sqrt{\delta^2+\Delta^2 \cos^2(\kappa d)}.
\ee
In the Bloch basis, the time independent Schr\"odinger equation with Hamiltonian (\ref{S}) reads
\bea
E\langle\chi_{0,\kappa}\ket{\Psi}&=&
E_{0,\kappa}\langle\chi_{0,\kappa}\ket{\Psi}+\rmi F \frac{\partial}{\partial \kappa}\langle\chi_{0,\kappa}\ket{\Psi}-Fd\bra{\chi_{0,\kappa}}\Psi\rangle-M_\kappa\bra{\chi_{1,\kappa}}\Psi\rangle \label{G2_64}\, \\
E\langle\chi_{1,\kappa}\ket{\Psi}&=&
E_{1,\kappa}\langle\chi_{1,\kappa}\ket{\Psi}+\rmi F \frac{\partial}{\partial \kappa}\langle\chi_{1,\kappa}\ket{\Psi}  +M_\kappa\bra{\chi_{0,\kappa}}\Psi\rangle \,, \label{G2_65}
\eea

\noindent where the coupling of the two bands is given by the transition matrix element
\be
  \bra{\chi_{1,\kappa'}}\hat H_{ZB}\ket{\chi_{0,\kappa}} =
  M_{\kappa}\delta_{\frac{\pi}{d}}(\kappa-\kappa') \label{deltafunc}
\ee
with the reduced matrix element
\bea
  M_{\kappa} &=& \rmi F \frac{\rka\frac{\partial}{\partial \kappa}u_\kappa\rkz v_\kappa- \rka\frac{\partial}{\partial \kappa}v_\kappa\rkz u_\kappa}{u_\kappa^2+v_\kappa^2}\,
  \rme^{\rmi   \kappa d}=   -\,\frac{\rmi F d \Delta \delta \sin(\kappa d)\,
  \rme^{\rmi \kappa d}}{2\delta^2+2\Delta^2\cos^2(\kappa d)} \,
  \label{G2_69}
\eea
(cf. \cite{Zhao91a}). The modulus of $M_\kappa$ is shown in figure \ref{2_2_a}. 

\noindent
Obviously, the band transitions mainly take place at the edge of the Brillouin zone and,
because of the delta-function in (\ref{deltafunc}), 
only direct interband transitions are possible.

\section{Spectral properties}
\label{sec-spectrum}

We now discuss the spectrum of the two-miniband Hamiltonian (\ref{S}) with an
external field.
The spectrum of the single-band Wannier-Stark Hamiltonian ($\delta=0$) consists
of a ladder of equally spaced eigenvalues -- the Wannier-Stark ladder.
Note that beyond the single-band tight-binding approximation these eigenstates
become resonances embedded into a continuum (see \cite{02wsrep} for a review).
In the case of two minibands, $\delta \ne 0$, the spectrum consists of
two ladders with an offset in between, as will be shown below.
Furthermore the dependence of the offset on the system parameters will
be discussed in detail. Certain aspects of the spectrum were previously
discussed in \cite{Zhao91a}.

To keep the notation simple, we introduce the translation operator
\be
\hat T_m=\sum\limits_{n=-\infty}\limits^{\infty}\ket{n-m}\bra{n} \,
\label{eqn-translation-op}
\ee
and an operator $\hat G$ that causes the inversion of the sign of $\delta$ in all following terms. They fulfil the following commutation relations
\bea
\eka\hat T_m,\hat H\ekz&=&\frac{\delta}{2}\sum\limits_{n=-\infty}\limits^{\infty} (-1)^n \eka1-(-1)^m\ekz \ket{n-m}\bra{n} \nn\\
&&\quad +Fd\sum\limits_{n=-\infty}\limits^{\infty} m\ket{n-m}\bra{n} \,, \\
\eka\hat G, \hat H\ekz&=&-\delta \sum\limits_{n=-\infty}\limits^{\infty} (-1)^n \ket{n}\bra{n} \, \hat G \quad\mbox{and}\quad\eka \hat T_m,\hat G\ekz=0 \, .
\eea
An eigenvector $\ket{\Psi}$ of $\hat H$ with the eigenvalue $E(\delta,\Delta,Fd)$,
\be
\hat H\ket{\Psi}=E(\delta,\Delta,Fd)\ket{\Psi} \, ,
\label{G2_102}
\ee
satisfies 
\bea
\hat H\gka \hat T_{2l}\ket{\Psi}\gkz&=&\gka E(\delta,\Delta,Fd)-2ldF \gkz\gka\hat T_{2l} \ket{\Psi}\gkz \, , \\
\hat H\gka \hat T_{2l+1}\hat G \ket{\Psi}\gkz&=& \gka E(-\delta,\Delta,Fd) -(2l+1)Fd \gkz\gka\hat T_{2l+1}\hat G \ket{\Psi}\gkz \,.
\eea
Thus the application of the operators $\hat T_{2l}$ resp. $\hat T_{2l+1}\hat G$
to the eigenvector $\ket{\Psi}$ yields two ladders of eigenvectors
\be
  \hat H \ket{\Psi_{\alpha,n}} = E_{\alpha,n}\ket{\Psi_{\alpha,n}} \quad\mbox{with}\quad \alpha=0, 1
  \label{eqn-eigeneqn-jn}
\ee
defined by
\be
  \ket{\Psi_{0,n}}=\hat T_{-2n}\ket{\Psi} \quad \mbox{and} \quad
  \ket{\Psi_{1,n}}=\hat G\hat T_{-(2n+1)}\ket{\Psi}
  \label{G2_107} \label{G2_121}
\ee
with equidistant eigenenergies
\bea
  E_{0,n}&=&E(\delta,\Delta,Fd)+2ndF \, ,\label{G2_120a}\\  
  E_{1,n}&=&E(-\delta,\Delta,Fd)+(2n+1)dF \, .
  \label{G2_120}
\eea
Thus the spectrum of the Hamiltonian (\ref{S}) consists of two energy ladders with equal spacings and
an offset given by $E(\delta,\Delta,Fd)$. Furthermore it can be shown that there are
not more than two different energy ladders (see appendix).
The eigenvectors $\ket{\Psi_{\alpha,n}}$ are displaced by $2n$ lattice periods
with respect to $\ket{\Psi_{\alpha,0}}$, where $\alpha$ labels the two ladders.

The dependence of $E(\delta,\Delta,Fd)$ on the system parameters $\delta$, $\Delta$
and $Fd$ shows several interesting symmetries. In the case $\delta=0$ it is well known
that for all values of $\Delta$ and $Fd$ we can choose $E(0,\Delta,Fd)=0$ \cite{04bloch1d}.
In the case $\delta\neq0$ it can be shown that
$E(-\delta,\Delta,Fd)=-E(\delta,\Delta,Fd)$ holds, as well as
$E(\delta,-\Delta,Fd)=E(\delta,\Delta,Fd)$. The proofs of these relations are
quite similar and only one of them is given here.
We consider the operator $\hat X$ defined by
\be
  \hat X=\sum\limits_{n=-\infty}\limits^{\infty}\ket{-n}\bra{n}\,(-1)^n \, ,
\ee
which fulfils the relation
\be
  \hat X \hat H(\delta, \Delta, Fd)=-\hat H(-\delta,\Delta, Fd) \hat X.
\ee
Applying $-\hat X$ to equation (\ref{eqn-eigeneqn-jn}) yields
\bea
  -E_{\alpha,n}(\delta, \Delta, Fd)\gka\hat X\ket{\Psi_{\alpha,n}} \gkz&=&
  \hat H(-\delta, \Delta, Fd)\gka\hat   X\ket{\Psi_{\alpha,n}} \gkz \nn\\
  &=:&E_{\alpha',n'}(-\delta, \Delta, Fd) \gka\hat X\ket{\Psi_{\alpha,n}} \gkz.
\eea
This means that the whole spectrum is antisymmetric by means of
\be
-E_{\alpha,n}(\delta, \Delta, Fd)=E_{\alpha',n'}(-\delta, \Delta, Fd) \,. \label{asym}
\ee
To simplify the following argumentation, we consider without loss of generality the whole
spectrum $\mathcal{S}$  modulo $2dF$. Thus we get
\be
\mathcal{S}=\lbrace E(\delta,\Delta,Fd), E(-\delta,\Delta,Fd)+dF \rbrace_{{\rm mod} \, 2dF} \, .
\ee
For $-\delta$ the spectrum therefore reads
\be
  \mathcal{S}=\lbrace E(-\delta,\Delta,Fd), E(\delta,\Delta,Fd)+
  dF \rbrace_{{\rm mod} \, 2dF} \, .\label{asym1}
\ee
By equation (\ref{asym}) this must be equal to
\bea
\mathcal{S}&=&\lbrace -E(\delta,\Delta,Fd), -E(-\delta,\Delta,Fd)-dF \rbrace_{{\rm mod} \, 2dF} \nn\\
&=&\lbrace -E(\delta,\Delta,Fd), -E(-\delta,\Delta,Fd)+dF \rbrace_{{\rm mod} \, 2dF} \, .\label{asym2}
\eea
From the equivalence of the equations (\ref{asym1}) and (\ref{asym2}) it follows that 
\be
E(-\delta,\Delta,Fd)=-E(\delta,\Delta,Fd)\, .
\ee

\begin{figure}[htb]
\centering
\includegraphics[height=7.6cm,  angle=0]{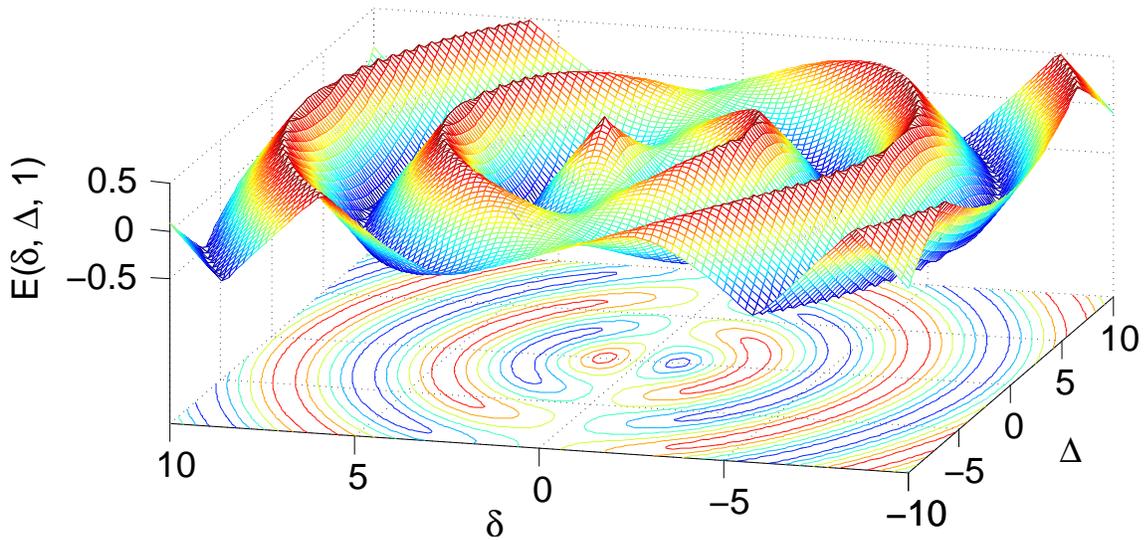}
\caption{\label{2_6} 
Alternating energy shift $E(\delta,\Delta,1)$ versus $\delta$ and $\Delta$ for $Fd=1$.}
\end{figure}

\begin{figure}[htb]
\centering
\includegraphics[height=5.7cm,  angle=0]{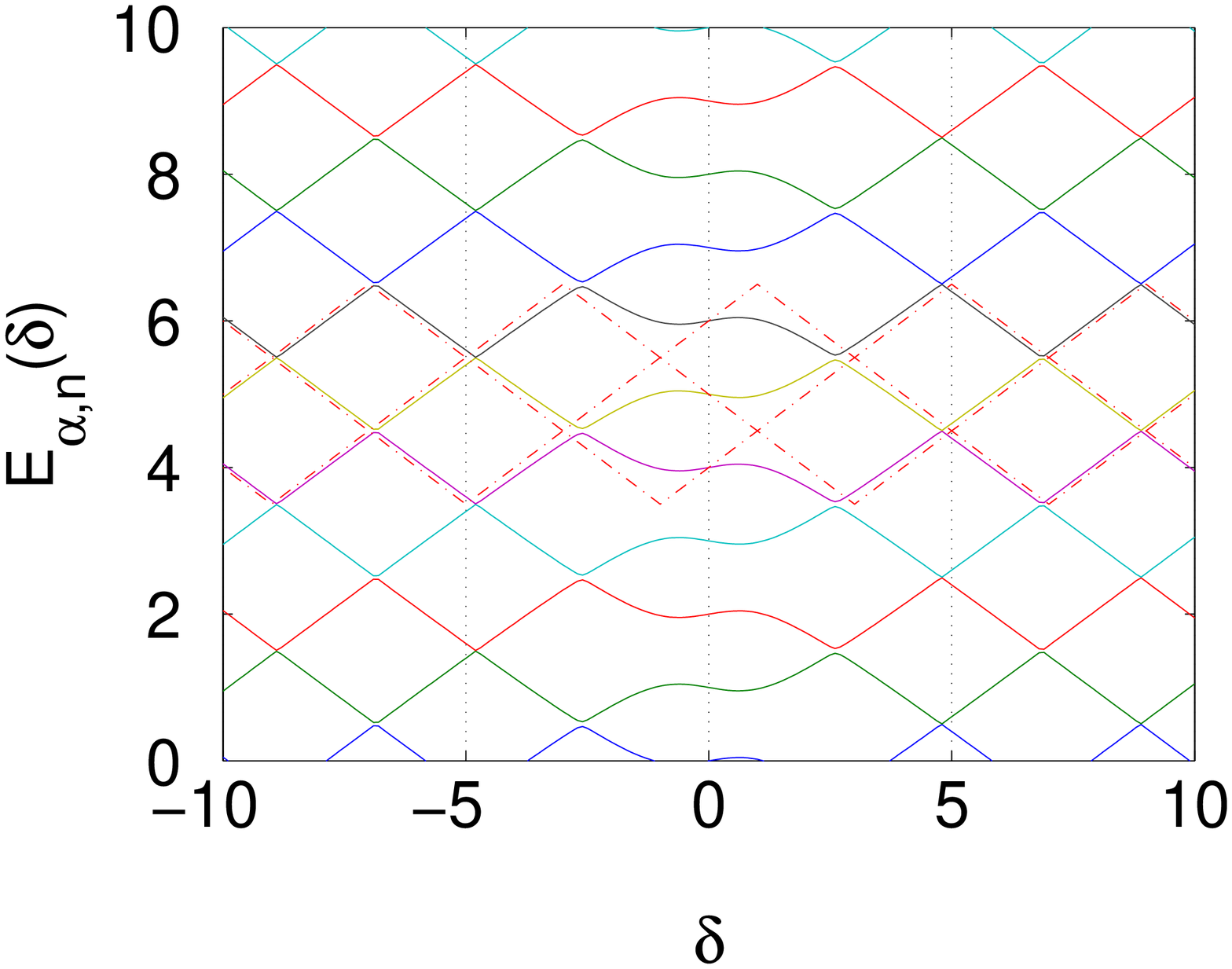}
\includegraphics[height=5.7cm,  angle=0]{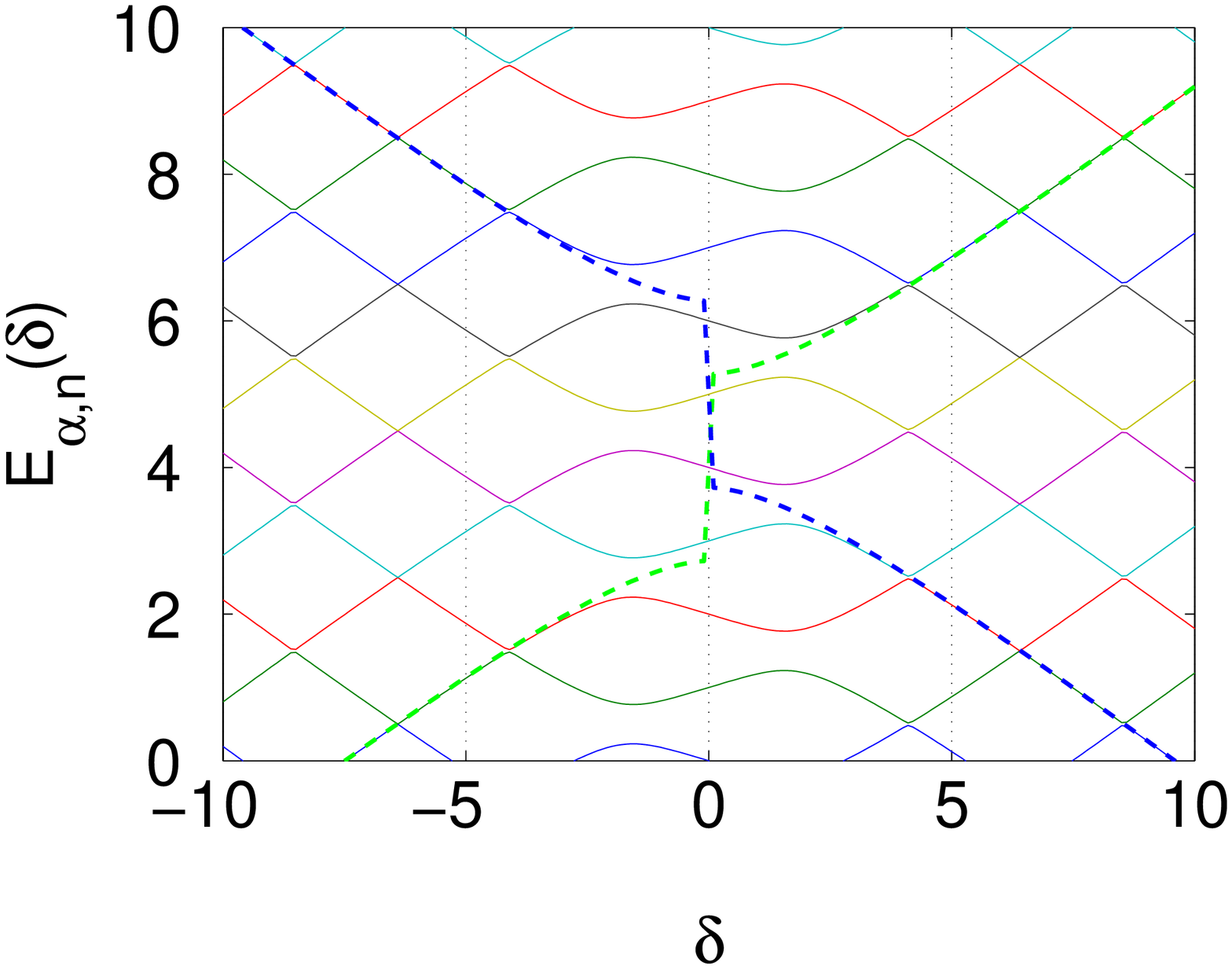}
\caption{\label{2_5} \label{3_1}
Spectrum $E_{\alpha,n}(\delta)$ of the Hamiltonian (\ref{S}) for $Fd=1$ (solid lines).
The left-hand side shows the case $\Delta=2$, where $\Delta=0$ is given as
reference (dashed line). Note the avoided crossings of the energy levels.
The right-hand side shows the spectrum for $\Delta=4$. Here two rungs of
different energy ladders calculated with the approximation $M_\kappa\approx0$
(dashed lines, cf. equation (\ref{eqn-offset-larged})) are shown for comparison.}
\end{figure}

Numerical results for the energy offset $E(\delta,\Delta,1)$ and the energy
ladders $E_{\alpha,n}(\delta,\Delta,1)$ are shown in figure
\ref{2_6} and figure \ref{3_1}.
In all examples we consider only the case $Fd=1$, as all other
cases can be reduced to it by a simple scaling
\be
  E_{\alpha,n}(\delta,\Delta,Fd)=
  dFE_{\alpha,n}\rka \textstyle\frac{\delta}{Fd}\displaystyle,\textstyle\frac{\Delta}{Fd},1\rkz\displaystyle \, ,
  \label{G2_135}
\ee
which directly follows from a rescaling of the Hamiltonian $\hat H$ by $Fd$.

In certain limits, one can easily derive some analytic approximations for
the ladder offset $E(\delta, \Delta, Fd)$:
\begin{enumerate}
\item For $\delta\gg\Delta$ the two minibands are well
separated energetically, so that their coupling can be neglected. Hence one sets
$M_\kappa\approx0$ in the equations (\ref{G2_64}) and (\ref{G2_65}).
The energy shift is then given by 
\bea
  E(\delta, \Delta, Fd)&=&\frac{\gamma}{\pi}\sqrt{\delta^2+\Delta^2}\,
  I\rka\textstyle\frac{\pi}{2}\displaystyle,\textstyle\sqrt{\frac{\Delta^2}{\delta^2+\Delta^2}}\displaystyle\rkz \\
  &=&\frac{\gamma}{\pi}\int_0^{\pi/2}\sqrt{\delta^2+\Delta^2\cos^2y}\, \rmd y \,
  \label{eqn-offset-larged}
\eea
(compare \cite{Zhao91a}) with $\gamma={\rm sgn}(\delta)$. Here, $I$ denotes the incomplete
elliptic integral of the second kind \cite{Abra72}.
This approximation is compared to the numerical results in figure \ref{3_1}.
\item For small $\delta$ perturbation theory can be done, which yields
\be
  E(\delta, \Delta, Fd)=\frac{\delta}{2}J_0\rka\textstyle\frac{\Delta}{Fd}\displaystyle\rkz \label{st}\, ,
  \label{eqn-offset-bessel}
\ee
where $J_0$ denotes the ordinary Bessel function.
\end{enumerate}

The zeros of $E(\delta,0,1)$ are $\delta=0,\pm2,\pm4,\ldots$ (see figure \ref{2_6}), whereas the zeros of $E(\delta,\Delta,1)$ for very small $\delta$ are given by the zeros of $J_0(\Delta)$ (see equation (\ref{st})) which are $\Delta\approx n\pi$ for large $\Delta$. These relations indicate that the characteristic structure of the energy shift looks similar for large $\Delta$ and large $\delta$.
In figure \ref{2_5} one clearly sees the asymmetry of $E(\delta,\Delta,1)$ 
with respect to $\delta$. Avoided crossings appear in the typical case of $\Delta\neq0$.

\section{Dynamics}
\label{sec-dynamics}
\subsection{An algebraic ansatz for the time evolution operator}

In this section we derive some general results for the time evolution operator
$\hat U$. The single-band system $\delta = 0$ can be treated conveniently within
a Lie-algebraic approach \cite{03TBalg}.
To simplify notation, we rewrite the Hamiltonian (\ref{S}) using the operators
\bea
  \hat N&=&\sum\limits_{n=-\infty}\limits^{\infty} n\ket{n}\bra{n}\, , \quad
  \hat L = \sum\limits_{n=-\infty}\limits^{\infty} (-1)^n\ket{n}\bra{n}=(-1)^{\hat N} \, ,\\
  \hat K&=&\sum\limits_{n=-\infty}\limits^{\infty} \ket{n-1}\bra{n} \quad \mbox{and} \quad
  \hat K^{\dagger} =\sum\limits_{n=-\infty}\limits^{\infty} \ket{n+1}\bra{n}\, .
\eea
These operators fulfil the commutator relations
\be
\eka \hat K, \hat N\ekz=\hat K\, , \quad\quad \eka \hat K^{\dagger}\, , \hat N\ekz=-\hat K^{\dagger}\, , \quad\quad \eka \hat K^{\dagger}, \hat K\ekz=0 \label{kom1}
\ee
and
\be
\eka \hat L, \hat N\ekz=0\, , \quad\quad
\eka \hat K^{\dagger}\, , \hat L\ekz=2\hat K^{\dagger}\hat L\, , \quad\quad
\eka \hat K, \hat L\ekz=2\hat K\hat L \, .
\label{kom}
\ee
From the commutator relations of $\hat K^{\dagger}$, $\hat K$ and $\hat L$ one
concludes that for any function $\hat f$ of the shift operators alone, one has
\be
\hat f(\hat K,\hat K^\dagger)\hat L=\hat L \hat f(-\hat K,-\hat K^\dagger) \, .
\ee
The Schr\"odinger equation for the time evolution operator $\hat U$ then reads
\be
  \rmi \hbar \frac{\partial}{\partial t}\hat U= \eka-\frac{\Delta}{4}\rka
  \hat K+\hat K^\dagger\rkz+\frac{\delta}{2}\hat L+Fd\hat N\ekz\hat U \, .
  \label{G4_7}
\ee
In analogy to the single-band system \cite{03TBalg} we make the ansatz
\be
\hat U =\rme^{-\rmi C(t)\hat N}\rka \hat A (\hat K,\hat K^\dagger,t) +\hat L \hat B(\hat K,\hat K^\dagger,t)\rkz
\label{G4_13}
\ee
where $C(t)$, $\hat A(\hat K,\hat K^\dagger,t)$ and $\hat B(\hat K,\hat K^\dagger,t)$
are yet unknown functions or operators.
To abbreviate the notation we do not write the arguments
of the operator valued functions $\hat A(t)$ and $\hat B(t)$ in the following explicitly.
Substituting equation (\ref{G4_13}) into (\ref{G4_7}) leads to
\bea
\hbar \,\dot{C}(t)\hat N \rme^{-\rmi C(t)\hat N}\rka \hat A\rka t\rkz+\hat L \hat B\rka t\rkz\rkz 
+\rmi \hbar \,\rme^{-\rmi C(t)\hat N}\rka \dot{\hat A}\rka t\rkz+\hat L \dot{\hat B}\rka t\rkz\rkz
\nonumber \\ 
\quad=Fd\hat N \rme^{-\rmi C(t)\hat N}\rka \hat A\rka t\rkz+\hat L \hat B \rka t\rkz\rkz 
-\frac{\Delta}{4}\hat K\rme^{-\rmi C(t)\hat N}\rka \hat A\rka t\rkz+\hat L \hat B\rka t\rkz\rkz \nn\\
\quad-\frac{\Delta}{4}\hat K^\dagger\rme^{-\rmi C(t)\hat N}\rka \hat A\rka t\rkz+\hat L \hat B\rka t\rkz\rkz 
+\frac{\delta}{2}\hat L \,\rme^{-\rmi C(t)\hat N}\rka \hat A\rka t\rkz+\hat L \hat B\rka t\rkz\rkz .
\label{G4_14} \label{G4_19}
\eea
Here and in the following, the dot marks the derivative with respect to the time $t$.
The first term of the left and the right side of equation (\ref{G4_14}) cancel
each other if
\be
  \hbar \, \dot C(t)=Fd \quad \Longrightarrow \quad
  C(t)=\frac{d}{\hbar}\int\limits_0^t F(\tau) \, \rmd \tau.
\ee
The remaining terms in equation (\ref{G4_14})
can be simplified by multiplying with $\rme^{\rmi C(t)\hat N}$ from the left
and using the relations
\bea
\rme^{\rmi C(t)\hat N}\hat K\rme^{-\rmi C(t)\hat N}&=&\rme^{-\rmi C(t)}\hat K \, ,\\
\rme^{\rmi C(t)\hat N}\hat K^\dagger\rme^{-\rmi C(t)\hat N}&=&\rme^{\rmi C(t)}\hat K^\dagger \,,
\eea
which follow from the Baker-Hausdorff formula
\be
\fl
\hspace{7mm}
\rme^{z\hat Y}\,\hat X\rme^{-z\hat Y}=\hat X+z\eka\hat Y, \hat X\ekz+\frac{z^2}{2!}\eka\hat Y, \eka\hat Y, \hat X\ekz \ekz
+\frac{z^3}{3!}\eka\hat Y, \eka\hat Y, \eka\hat Y, \hat X\ekz\ekz \ekz+ \ldots .
\ee
Since $\hat L$ and $\hat N$ commute and $\hat L \hat L=\eye$,
one finally arrives at
\bea
&&\hat L\rka \rmi \hbar \dot{\hat B}\rka t\rkz
              -\frac{\Delta}{4} \rme^{-\rmi C(t)}\hat K \hat B \rka t\rkz
              -\frac{\Delta}{4} \rme^{\rmi C(t)}\hat K^\dagger \hat B\rka t\rkz
              -\frac{\delta}{2} \hat A\rka t\rkz \rkz \nn\\
&& \quad =-\rka \rmi \hbar \dot{\hat A} \rka t\rkz
        +\frac{\Delta}{4} \rme^{-\rmi C(t)}\hat K \hat A\rka t\rkz
        +\frac{\Delta}{4} \rme^{\rmi C(t)}\hat K^\dagger \hat A\rka t\rkz
        -\frac{\delta}{2} \hat B\rka t\rkz \rkz \,.
\eea
Obviously, this equation is fulfiled if each bracket is zero. The result are two coupled differential equations
\bea
\dot{\hat A}\rka t\rkz&=&
\rmi\frac{\Delta}{4\hbar} \rka \rme^{-\rmi C(t)}\hat K + \rme^{\rmi C(t)}\hat K^\dagger\rkz \hat A\rka t\rkz -\rmi\frac{\delta}{2\hbar} \hat B\rka t\rkz \label{G4_39}\, ,\\
\dot{\hat B}\rka t\rkz&=&
-\rmi\frac{\Delta}{4\hbar} \rka \rme^{-\rmi C(t)}\hat K + \rme^{\rmi C(t)}\hat K^\dagger\rkz \hat B\rka t\rkz -\rmi\frac{\delta}{2\hbar} \hat A\rka t\rkz \, , \label{G4_40}
\eea
that only depend on the commuting operators $\hat K$ and $\hat K^\dagger$.
With the initial condition $\hat U(0)=\eye$ and the equations (\ref{G4_13}), (\ref{G4_39}) and (\ref{G4_40}) we conclude
\bea
\hat A\rka 0\rkz=\eye \,, \quad \quad
&\dot{\hat A}\rka 0\rkz=\rmi\frac{\Delta}{4\hbar} \rka \hat K + \hat K^\dagger\rkz \, ,  \\
\hat B\rka 0\rkz=0 \,, \quad \quad
&\dot{\hat B}\rka 0\rkz=-\rmi\frac{\delta}{2\hbar} \, .
\eea

Without loss of generality, we continue with the representation of $\hat K$ and $\hat K^\dagger$
in the $\kappa$-basis of the single-band system
\be
|\kappa \rangle =\sqrt{\frac{d}{2 \pi }} \sum\limits_{n=-\infty}\limits^{\infty} \rme^{\rmi n \kappa d} |n \rangle \,,
\ee
where the operators are diagonal with respect to the quasi-momentum $\kappa$:
\be
\bra{\kappa'}\hat K\ket{\kappa}=\rme^{\rmi \kappa d}\delta_{\frac{2\pi}{d}}(\kappa-\kappa') \quad \mbox{and} \quad \bra{\kappa'}\hat K^\dagger\ket{\kappa}=\rme^{-\rmi \kappa d}\delta_{\frac{2\pi}{d}}(\kappa-\kappa')
\ee
with the $\frac{2\pi}{d}$-periodic delta function $\delta_{\frac{2\pi}{d}}$.
The operators $\hat A(\hat K,\hat K^\dagger,t)$ and $\hat B(\hat K,\hat K^\dagger,t)$
in the $\kappa$-basis are denoted by $A_\kappa(t)$ and $B_\kappa(t)$ which leads to the following system
of ordinary differential equations
\bea
\dot A_\kappa \rka t\rkz&=&
\rmi\frac{\Delta}{2\hbar} \,\cos\rka \kappa d-C(t)\rkz A_\kappa \rka t\rkz
-\rmi\frac{\delta}{2\hbar} B_\kappa \rka t\rkz \, , \label{G4_43}\\
\dot B_\kappa \rka t\rkz&=&
-\rmi\frac{\Delta}{2\hbar} \,\cos\rka \kappa d-C(t)\rkz B_\kappa \rka t\rkz
-\rmi\frac{\delta}{2\hbar} A_\kappa \rka t\rkz \, . \label{G4_44}
\eea
From now on we consider the case of a constant force $C(t)=\frac{dFt}{\hbar}$. By substituting
\be
x=\kappa d-\frac{dFt}{\hbar} \quad \mbox{and} \quad \frac{\partial}{\partial t}=-\frac{dF}{\hbar}\frac{\partial}{\partial x}
\ee
one obtains
\bea
A'_\kappa (x)&=&
-\rmi\frac{\Delta}{2dF} \,\cos(x) A_\kappa(x)
+\rmi\frac{\delta}{2dF} B_\kappa(x) \, ,
\label{G4_46}\\
B'_\kappa(x)&=&
\rmi\frac{\Delta}{2dF} \,\cos(x) B_\kappa(x)
+\rmi\frac{\delta}{2dF} A_\kappa(x) \, ,
\label{G4_47}
\eea
where the prime denotes derivation with respect to $x$.
Differentiating again with respect to $x$
decouples these equations with the result
\bea
\fl \qquad
A''_\kappa(x)+\rka\rka\textstyle\frac{\delta}{2dF}\displaystyle\rkz^2-\rmi \rka\textstyle\frac{\Delta}{2dF}\displaystyle\rkz \sin x +\rka\textstyle\frac{\Delta}{2dF}\displaystyle\rkz^2 \cos^2 x \rkz A_\kappa(x)&=&0 \label{G4_50} \, ,\\
\fl \qquad
B''_\kappa(x)+\rka\rka\textstyle\frac{\delta}{2dF}\displaystyle\rkz^2+\rmi \rka\textstyle\frac{\Delta}{2dF}\displaystyle\rkz \sin x +\rka\textstyle\frac{\Delta}{2dF}\displaystyle\rkz^2 \cos^2 x \rkz B_\kappa(x)&=&0 \label{G4_53} \, .
\eea
Both equations are of the type of Whittaker-Hill differential equations
as described, e.g., in \cite{Urwi70}. A solution of the equations (\ref{G4_50}), (\ref{G4_53})
in closed form is only known in the cases $\delta=0$ or $\Delta=0$. Needless to say that the
equations (\ref{G4_43}) and (\ref{G4_44}) can be solved in the case $F=0$.
There always exists a solution of the differential equations (\ref{G4_50}), (\ref{G4_53})
which justifies the ansatz (\ref{G4_13}), even if we can not construct such a solution in
closed form.

\subsection{Time evolution}
A wave packet, whose dynamics is governed by the double-periodic Hamiltonian (\ref{S}) is expected to
show both, Bloch oscillations and Zener tunneling simultaneously.
The tunneling rate at the edge of the Brillouin zone is approximately given by the
famous Landau-Zener formula \cite{Land32, Zene32}
\be
  P\approx\rme^{-\frac{\pi\delta^2}{2d|F\Delta|}} \label{lzf} \, .
\ee
In this chapter we study Bloch-Zener oscillations, a coherent superposition of Bloch
oscillations and Zener tunneling.
Let $\ket{\Psi_{\alpha,n}}$ be the eigenfunctions of the Hamiltonian
(\ref{S}), called Wannier-Stark functions, as introduced in section
\ref{sec-spectrum}. The index $\alpha=0,1$ labels the two energy ladders.
We expand the Wannier-Stark functions in the $\kappa$-basis,
\be
\ket{\Psi_{\alpha,n}}=\int\limits_{-\frac{\pi}{2d}}^{\frac{\pi}{2d}}a_{\alpha,n}(\kappa)\ket{\chi_{0,\kappa}}\, \rmd \kappa
+\int\limits_{-\frac{\pi}{2d}}^{\frac{\pi}{2d}}b_{\alpha,n}(\kappa)\ket{\chi_{1,\kappa}}\, \rmd \kappa 
\label{eqn-WSfunct-kappa1}
\ee
with unknown coefficients $a_{\alpha,n}(\kappa)$ and $b_{\alpha,n}(\kappa)$.
According to equation (\ref{G2_107}), the Wannier-Stark functions with
different site indices $n$ are related by the translation operator $\hat T_m$
(see equation (\ref{eqn-translation-op})).
The Bloch waves of the two band system defined by the equations
(\ref{G2_18}), (\ref{G2_19}) obey the relations
\be
  \hat T_{-2n}\ket{\chi_{\beta,\kappa}}=\rme^{-\rmi2n\kappa d}\ket{\chi_{\beta,\kappa}} \,.
\ee
Thus the Wannier-Stark functions (\ref{eqn-WSfunct-kappa1}) can be written as
\be
\ket{\Psi_{\alpha,n}}=\int\limits_{-\frac{\pi}{2d}}^{\frac{\pi}{2d}}a_{\alpha,0}(\kappa)\rme^{-\rmi2n\kappa d}\ket{\chi_{0,\kappa}}\, \rmd \kappa
+\int\limits_{-\frac{\pi}{2d}}^{\frac{\pi}{2d}}b_{\alpha,0}(\kappa)\rme^{-\rmi2n\kappa d}\ket{\chi_{1,\kappa}}\, \rmd \kappa
\ee
with projections onto the Bloch states
\be
\bra{\chi_{0,\kappa}}\Psi_{\alpha,n}\rangle=a_{\alpha,0}(\kappa)\,\rme^{-\rmi2n\kappa d} \quad\mbox{and}\quad
\bra{\chi_{1,\kappa}}\Psi_{\alpha,n}\rangle=b_{\alpha,0}(\kappa)\,\rme^{-\rmi2n\kappa d}.
\ee
The time evolution of an arbitrary initial state $\ket{\psi(0)}$ can now be calculated
in the Wannier-Stark basis.
Expanding the initial state in the Wannier-Stark basis,
\be
  \ket{\psi(0)}=\sum\limits_n  c_{0,n} \ket{\Psi_{0,n}}+\sum\limits_n c_{1,n} \ket{\Psi_{1,n}} \, ,
\ee
the dynamics of $\ket{\psi}$ is given by
\be
  \ket{\psi(t)}=\sum\limits_n c_{0,n} \, \rme^{-\frac{\rmi}{\hbar}E_{0,n}t} \ket{\Psi_{0,n}}
  +\sum\limits_n c_{1,n} \, \rme^{-\frac{\rmi}{\hbar}E_{1,n}t} \ket{\Psi_{1,n}} \,.
\ee
The contributions of the Bloch bands are obtained by projecting onto
$\ket{\chi_{\alpha,\kappa}}$, which yields
\bea
\langle \chi_{0,\kappa}\ket{\Psi(t)}&=&\sum\limits_n c_{0,n} \, \rme^{-\frac{\rmi}{\hbar}(2ndF+E_0)t} a_{0,0}(\kappa)\,\rme^{-\rmi2n\kappa d}\nn\\
&&\quad+\sum\limits_n c_{1,n} \, \rme^{-\frac{\rmi}{\hbar}(2ndF+dF-E_0)t} a_{1,0}(\kappa)\,\rme^{-\rmi2n\kappa d} \, ,\\
\langle \chi_{1,\kappa}\ket{\Psi(t)}&=&\sum\limits_n c_{0,n} \, \rme^{-\frac{\rmi}{\hbar}(2ndF+E_0)t} b_{0,0}(\kappa)\,\rme^{-\rmi2n\kappa d}\nn\\
&&\quad+\sum\limits_n c_{1,n} \, \rme^{-\frac{\rmi}{\hbar}(2ndF+dF-E_0)t} b_{1,0}(\kappa)\,\rme^{-\rmi2n\kappa d} \, ,
\eea
where the eigenenergies (\ref{G2_120a}), (\ref{G2_120}) have been inserted and $E(\delta,\Delta,Fd)$ has been denoted by $E_0$.
One observes that two Fourier series appear for the two ladders
$\alpha=0,1$,
\be
   C_\alpha\rka\textstyle\kappa+\frac{Ft}{\hbar}\displaystyle\rkz = \sum\limits_n c_{\alpha,n}\,
   \rme^{-\rmi 2n d   \rka\kappa+\frac{Ft}{\hbar}\rkz} \, ,
\ee
which are $\pi/d$-periodic. Thus the time evolution can be written as
\bea
\fl \quad &&\langle \chi_{0,\kappa}\ket{\Psi(t)} =
  \rme^{-\frac{\rmi}{\hbar}E_0t}\eka a_{0,0}(\kappa)\,C_0\rka\textstyle\kappa+\frac{Ft}{\hbar}\displaystyle\rkz
  +a_{1,0}(\kappa)\,\rme^{-\frac{\rmi}{\hbar}(dF-2E_0)t}C_1\rka\textstyle\kappa+\frac{Ft}{\hbar}\displaystyle\rkz\ekz
  \label{G5_31} \, ,\\
\fl \quad && \langle \chi_{1,\kappa}\ket{\Psi(t)} =
  \rme^{-\frac{\rmi}{\hbar}E_0t}\eka b_{0,0}(\kappa)\,C_0\rka\textstyle\kappa+\frac{Ft}{\hbar}\displaystyle\rkz
  +b_{1,0}(\kappa)\,\rme^{-\frac{\rmi}{\hbar}(dF-2E_0)t}C_1\rka\textstyle\kappa+\frac{Ft}{\hbar}\displaystyle\rkz\ekz \, .
  \label{G5_32}
\eea
This structure of the dynamics causes some interesting effects discussed in the
following section.
\subsection{Bloch-Zener oscillations and reconstruction}

The dynamics of the two band system is characterized by two periods. The functions $C_0$ and $C_1$ are reconstructed at multiples of
\be
T_1=\frac{\pi \hbar}{d F} \label{t1} \, ,
\ee
whereas the exponential function $\rme^{-\frac{\rmi}{\hbar}(dF-2E_0)t}$ has a period of
\be
T_2=\frac{2\pi \hbar}{dF-2E_0} \, . \label{t2}
\ee
The period $T_1$ is just half of the Bloch time $T_B=\frac{2\pi \hbar}{d F}$ for the single-band system, $\delta=0$, which reflects the double periodicity of the two-band model. In the case of a wave function confined to a single energy ladder, one of the functions $C_0$ and $C_1$ is zero for all times $t$ and the initial state is reconstructed up to a global phase after a period $T_1$, which is just an ordinary Bloch oscillation.

In general, the functions (\ref{G5_31}) and (\ref{G5_32}) reconstruct up to a global phase
if the two periods $T_1$ and $T_2$ are commensurable,
\be
  \frac{T_2}{T_1}=\frac{2dF}{dF-2E_0}=\frac{r}{s} \quad\mbox{with}\quad r,s\in\mathbb{N}  \label{ratio}.
\ee
The reconstruction takes place at integer multiples of the Bloch-Zener time
\be
  T_{BZ}=sT_2=rT_1.
\ee
Examples of such reconstructing Bloch-Zener oscillations are shown in the next section.
Furthermore we can calculate the dynamics of the occupation probability at
multiples of $T_1$ by the aid of the equations (\ref{G5_31}), (\ref{G5_32}),
which yields
\bea
&&\int\limits_{-\frac{\pi}{2d}}^{\frac{\pi}{2d}}|\langle \chi_{0,\kappa}\ket{\Psi(nT_1)}|^2\,\rmd\kappa=X+Y\cos\rka\frac{dF-2E_0}{dF}\pi\, n+\varphi\rkz \label{G5_71} \, ,\\
&&\int\limits_{-\frac{\pi}{2d}}^{\frac{\pi}{2d}}|\langle \chi_{1,\kappa}\ket{\Psi(nT_1)}|^2\,\rmd\kappa=
1-\eka X+Y\cos\rka\frac{dF-2E_0}{dF}\pi\, n+\varphi\rkz\ekz , \label{G5_72}
\eea
where $X$ and $Y$ are real positive numbers. In the commensurable case (\ref{ratio}) we find
\bea
&&\int\limits_{-\frac{\pi}{2d}}^{\frac{\pi}{2d}}|\langle \chi_{0,\kappa}\ket{\Psi(nT_1)}|^2\,\rmd\kappa=X+Y\cos\rka2\pi\,\frac{s}{r} n+\varphi\rkz
\label{G5_74} \, ,\\
&&\int\limits_{-\frac{\pi}{2d}}^{\frac{\pi}{2d}}|\langle \chi_{1,\kappa}\ket{\Psi(nT_1)}|^2\,\rmd\kappa=
1-\eka X+Y\cos\rka2\pi\,\frac{s}{r} n+\varphi\rkz\ekz \, ,
\label{G5_75}
\eea 
where one recognizes the reconstruction at integer multiples of $T_{BZ}$.
In the case of incommensurable periods $T_2$ and $T_1$, we obtain $\varphi=0$
whenever only one of the bands is occupied at the beginning. Otherwise one
of the occupation probabilities would become greater than one for some $n$ whereas the
other one would become negative.

\subsection{Oscillating and breathing modes}

\begin{figure}[htb]
\centering
\includegraphics[height=7.9cm,  angle=0]{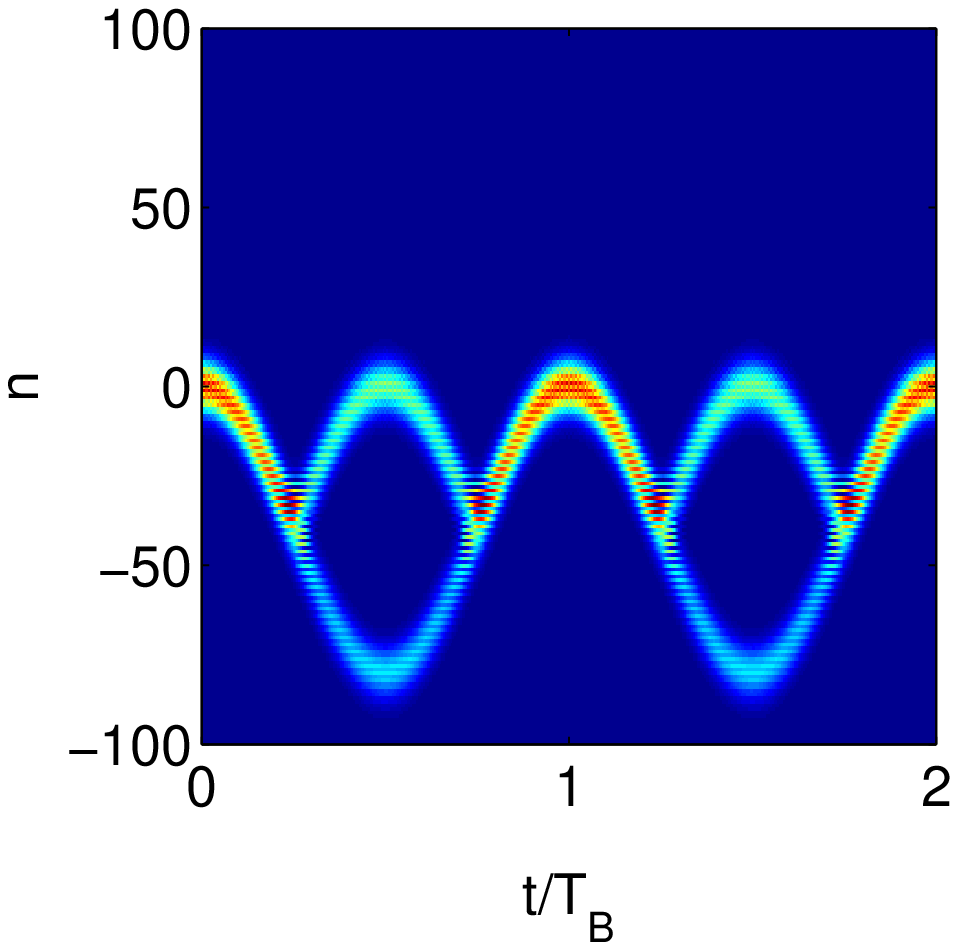}
\includegraphics[height=7.9cm,  angle=0]{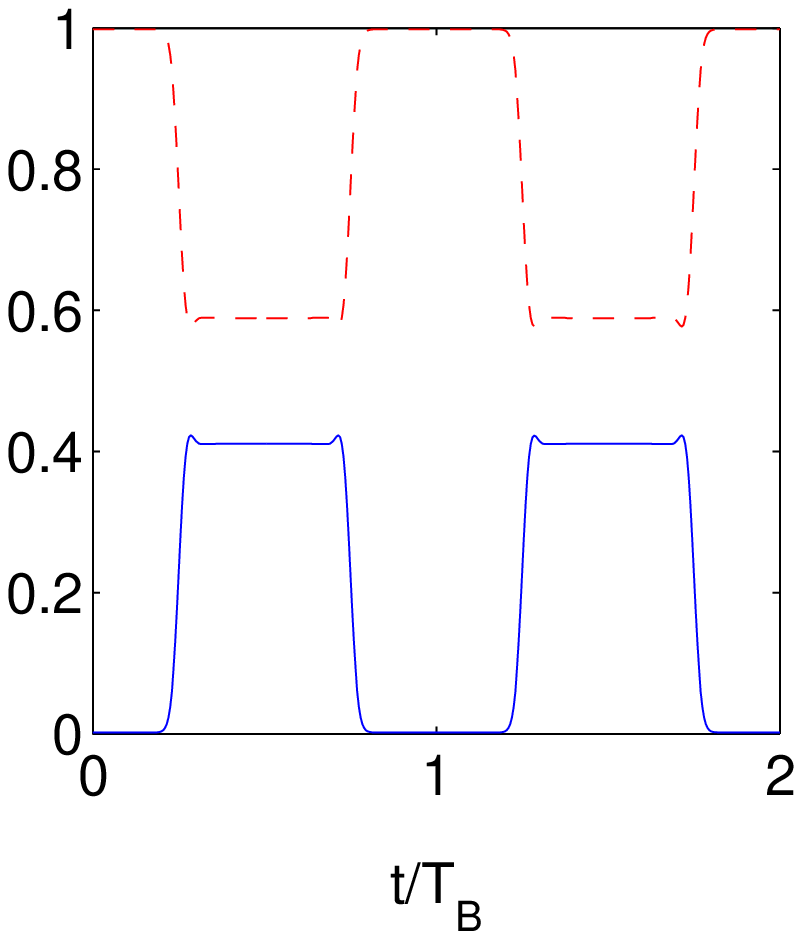}
\caption{\label{5_1}
Dynamics of an oscillating Bloch-Zener mode. The left-hand side shows $|\psi|^2$
versus 'space' and time.
The right-hand side shows the occupation probability of the
lower minband (dashed red line) and the upper band (solid blue line)
versus time. The parameters are $\Delta=80$, $dF=1$ and $\delta=6.734$.}
\end{figure}

The reconstructing Bloch-Zener oscillations discussed in the preceding section
will be illustrated by some numerical calculations. The dynamics of a so-called oscillatory
mode is shown in figure \ref{5_1}. Because of the strong localization of the Wannier
functions, this resembles closely the behaviour in real space. The initial state is chosen
as a broad gaussian distribution in the Wannier representation, namely
$\sim\rme^{-n^2/100}$.
By the choice of the parameters we have $E_0=0$ and we obtain a reconstruction
at integer multiples of the Bloch time $T_{BZ}=T_B$. Here and in the following, we take the
Bloch time $T_B=\frac{2\pi\hbar}{dF}=2T_1$ of the single-band system $\delta=0$
as the reference time scale.
Needless to say that many other reconstruction times $T_{BZ}$ can be adjusted in this way.
The edge of the Brillouin zone is achieved at multiples of $T_1=T_B/2$, where
the Bloch bands are close together and the probability for a band transition
reaches a maximum.

\begin{figure}[htb]
\centering
\includegraphics[height=6.8cm,  angle=0]{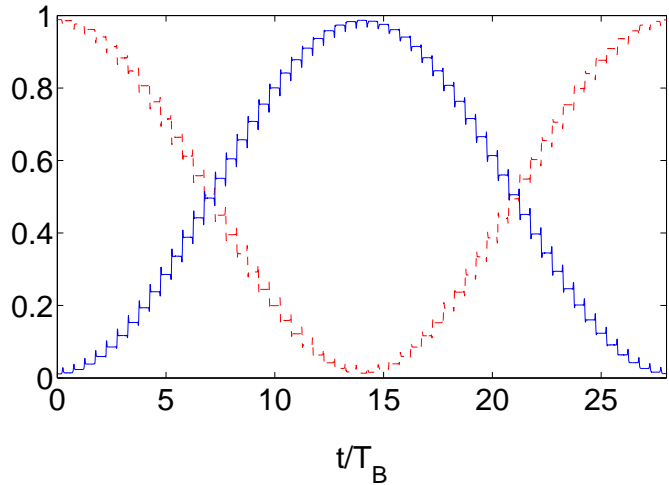}
\caption{\label{5_3}
Occupation probability of the lower minband (dashed red line) and the upper band
(solid blue line) for an oscillatory mode versus time.
The parameters are $\Delta=80$, $dF=1$ and $\delta=17.19$.}
\end{figure}

When $\frac{dF-2E_0}{dF}\pi$ is close to $2\pi$,
the cosine in equations (\ref{G5_74}) and (\ref{G5_75}) changes very slightly after
each period $T_1$, which illustrates the cosine-like behaviour of the occupation probability
very clearly. This is shown in figure \ref{5_3} for an oscillatory mode with the same initial
condition as above.
By the choice of the parameters $\Delta$, $dF$ and $\delta$ we get $E_0=-0.5178$ and hence
\be
\frac{T_2}{T_1}=\frac{56}{57} \, .
\ee
Thus we get a reconstruction at multiples of $T_{BZ}=28T_B$. The slowly varying oscillation
of the occupation between the bloch bands is similar to the Rabi oscillations for a two level
system.

\begin{figure}[htb]
\centering
\includegraphics[height=6.8cm,  angle=0]{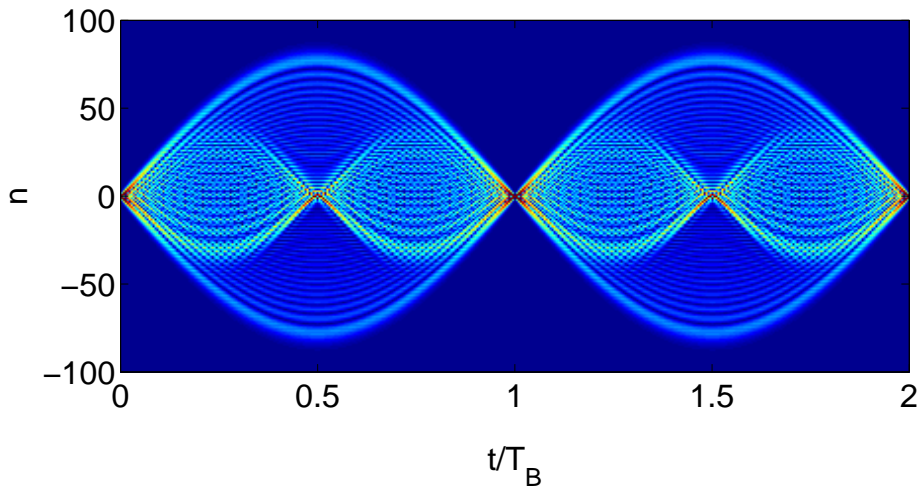}
\caption{\label{5_4}
Dynamics of a breathing mode. The modulus of the wave function $|\psi|$ is plotted 
versus 'space' and time.  The parameters are $\Delta=80$, $dF=1$ and $\delta=6.734$.}
\end{figure}

Let us now consider a wave packet that is initially located at 
a single site (see figure \ref{5_4}). In the case of a single-band system ($\delta=0$) one finds a
breathing behaviour: The width of the wave packet oscillates strongly, while its
position remains constant \cite{04bloch1d}. The parameters are chosen as in figure
\ref{5_1}, so that reconstruction happens again after integer multiples of $T_{BZ}=T_B$.
One observes that the enveloping structure with the period $T_{BZ}=T_B$ is overlayed
by a breathing mode of smaller amplitude, which can be interpreted as a fraction
of the wave packet staying in one band all the time. Therefore the characteristic period
of this part is $T_1=T_B/2$. The gradient of the dispersion relation of a band changes
at the edge of the Brillouin zone if there is no band transition. Therefore the direction
of motion given by the group velocity changes after $T_1=T_B/2$ and the breathing amplitude
of this part of the wave packet is thus smaller than the amplitude of the complete
breathing mode.

\subsection{Beam splitters}

The Bloch-Zener transition between the two minibands can be used
to construct a matter wave beam splitter with controllable 'transmission'.
To see this, note that interband transitions mainly take place at the edge of the
Brillouin zone (see figure \ref{2_2_a}). 
In the case of a partial band transition, the wave packet is spatially
separated, as illustrated in figure \ref{5_1}.

\begin{figure}[htb]
\centering
\includegraphics[height=7.2cm,  angle=0]{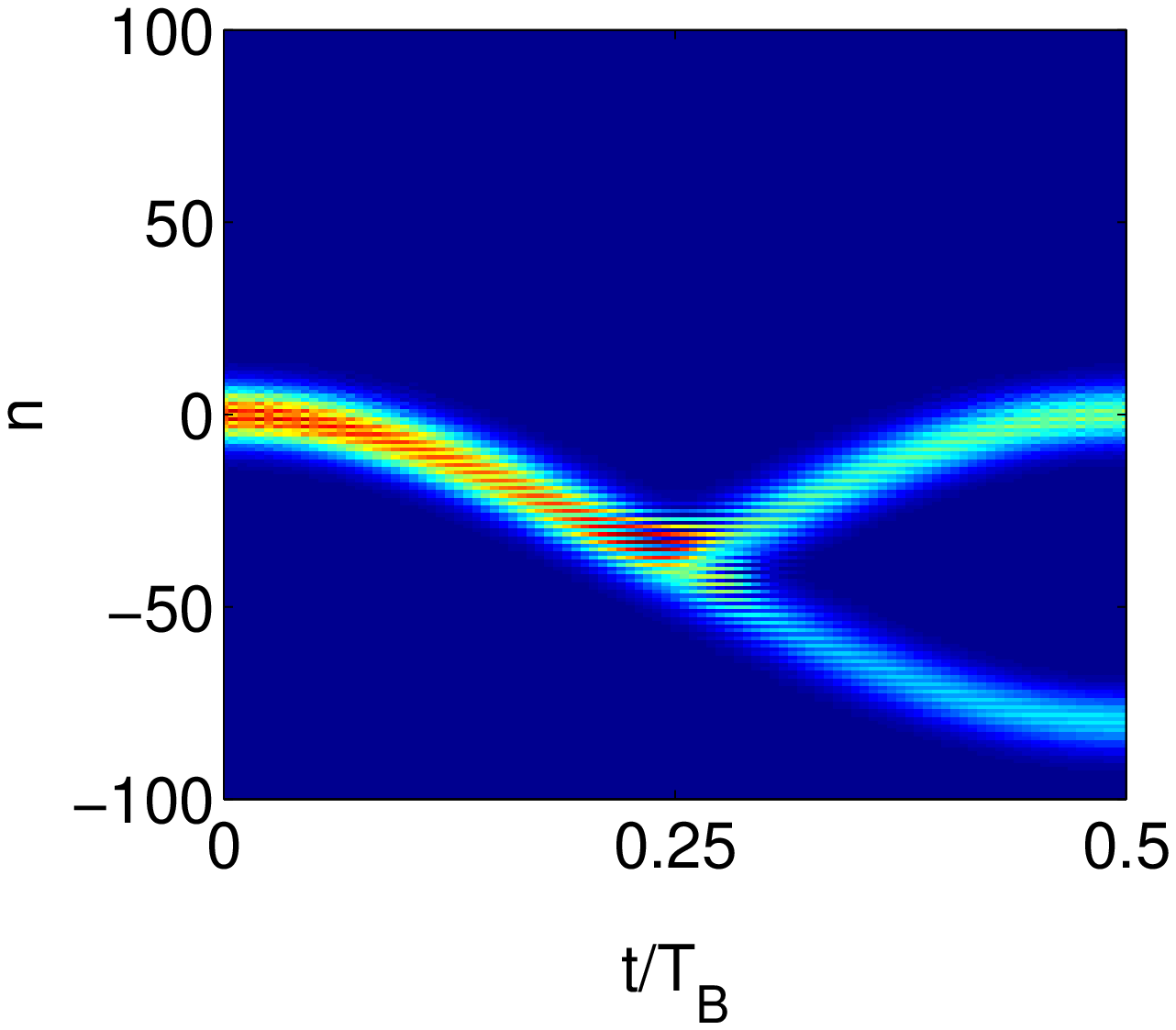}
\includegraphics[height=7.2cm,  angle=0]{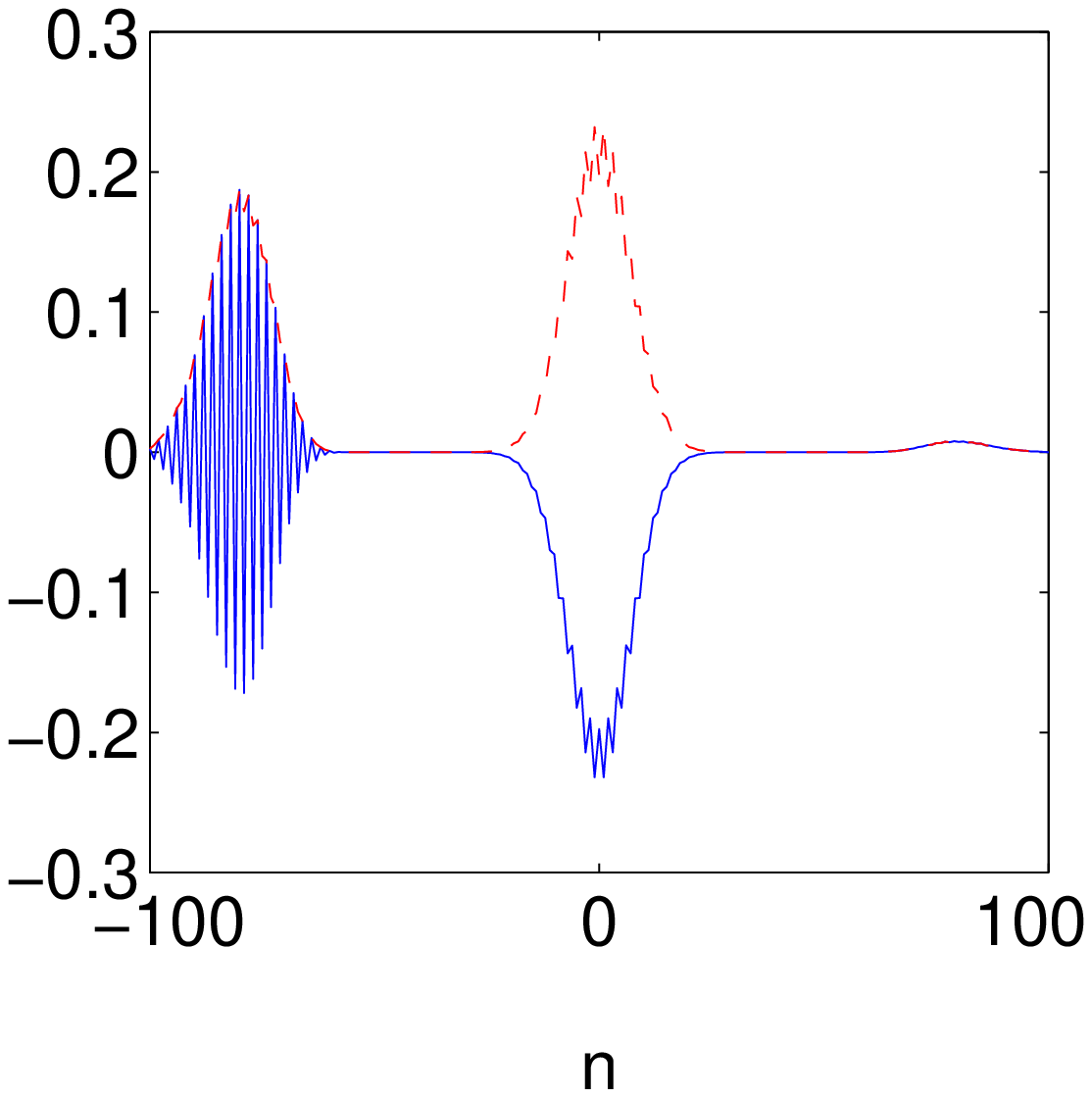}
\caption{\label{5_9}
Splitting of a gaussian wave packet in 'real' space. The left hand side shows 
$|\psi|^2$ versus 'space' and time. The right hand side shows the absolute value 
(red dashed line) and the real part (blue solid line) of the wave function at $t=0.5\,T_B$ 
versus 'space'. The parameters are $\Delta=80$, $dF=1$ and $\delta=6.734$.}
\end{figure}

\begin{figure}[htb]
\centering
\includegraphics[height=6.9cm,  angle=0]{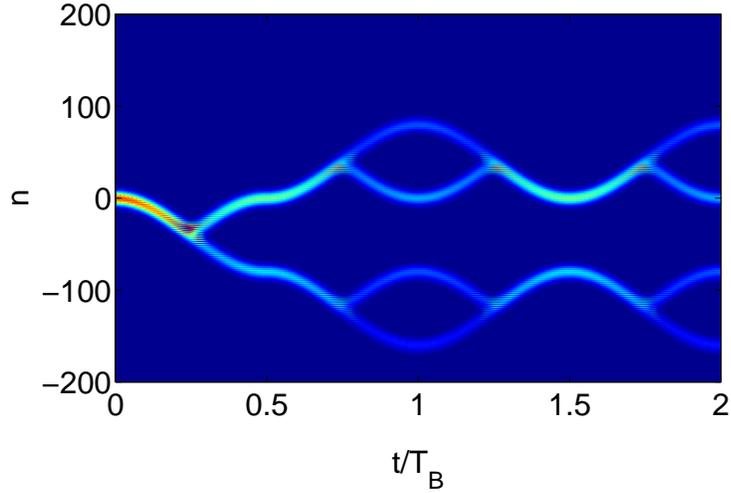}
\caption{\label{5_11}
Splitting of a gaussian wave packet in 'real' space. Shown is $|\psi|^2$ versus 
'space' $n$ and time $t$. The parameters are $\Delta=80$, $d=1$, $\delta=6.734$ and $|F|=1$.}
\end{figure}

The time evolution of the squared modulus of a wave packet with a broad initial
distribution $\sim\rme^{-n^2/100}$ is shown in figure \ref{5_9}.
At time $t=0.5\,T_B$, both parts of the wave packet have a gaussian profile, but
the lower part of the wave function consists of states with opposite momentum.
Hence the phase of the wave function and therefore its real part oscillate strongly.
Usually both parts merge again after a full Bloch period $T_B$ (cf. figure \ref{5_1}).
However, a permanent splitting of the wave packet can be obtained by flipping the sign
of $F$ at $t=0.5\,T_B$.
Thereby both parts of the wave packet are transported in opposite directions (compare \cite{06manipulation}). This effect
is shown in figure \ref{5_11} (same initial state as above). After flipping the field $F$, the
external force is constant and therefore both splitted wave packets continue their
Bloch-Zener oscillations separately. Furthermore it is possible to control the occupation 
of each branch by the choice of $\delta$.

\begin{figure}[htb]
\centering
\includegraphics[height=5.3cm,  angle=0]{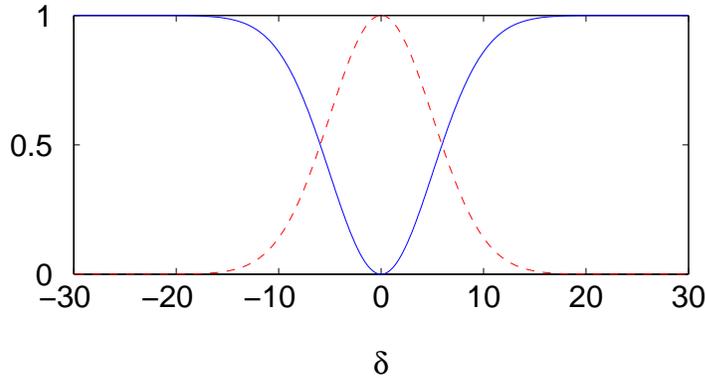}
\caption{\label{5_12}
Occupation probability of the above gaussian wave packet at $t=0.5\,T_B$ versus $\delta$. 
The red dashed line shows the occupation in the interval $-100\leq n\leq-40$ whereas the 
blue solid line shows the occupation in the interval $-41\leq n\leq20$. The parameters 
are $\Delta=80$, $d=1$, and $F=1$.}
\end{figure}

The transition probability at $t=0.5\,T_B$ for a wave packet initially located in a single band 
is in good approximation given by the Landau-Zener formula (\ref{lzf}). The numerical results for the 
occupation of the different branches (compare figure \ref{5_9}) at $t=0.5\,T_B$ versus $\delta$ 
are shown in figure \ref{5_12}. Such a controlled manipulation of the band gap can be achieved, 
for example, in optical lattices. In this case it is also possible to construct a beam splitter 
in a different way, namely by 'switching off' $\delta$ completely at $t=0.5\,T_B$. Instead of 
Bloch-Zener oscillations, simple Bloch oscillations will appear with the advantage that the 
two wave packets can be separated by low-loss transport as described in 
\cite{04bloch1d,06manipulation}. In any case it is possible to control the amplitude of each 
of the splitted wave packets by the choice of $\delta$.

\section{Conclusion and Outlook}
In conclusion, we have investigated the interplay between Bloch
oscillations and Zener tunneling between the Bloch bands.
To this end we considered a special tight-binding model under the
influence of a static external field. 
This model satisfactorily describes the dynamics of a more realistic 
system as demonstrated in a comparison with exact numerical computations \cite{06manipulation}.
An additional period-doubled potential leads
to a splitting of the ground band into two minibands in the field-free
case. For a non-zero external field, the spectrum of the system consists of
exactly two Wannier-Stark ladders. The dependence of the energy offset between
the two ladders on the system parameters was discussed in detail.

The dynamics of the system is governed by two time-scales, the Bloch
period $T_1$ defined in equation (\ref{t1}) and the period $T_2$ of the Zener
oscillation between the two energy ladders introduced in
equation (\ref{t2}).
If the two periods are commensurable, a wave packet will reconstruct
periodically up to a global phase. The occupation of the two minibands at multiples
of the Bloch period $T_1$ oscillates sinusoidally in time.
Furthermore it was shown that the dynamics of the system can be reduced
to a Whittaker-Hill differential equation. Unfortunately few analytic
results are available for the solutions of this differential equation.

Finally some numerical examples of Bloch-Zener oscillations were given.
It was shown that Bloch-Zener oscillations in a double-periodic
potential can be used to construct matter wave beam splitters, e.g. for cold atoms
in optical lattices.

A further important application of a Bloch-Zener oscillation as shown in
figure \ref{5_1} is the construction of a matter wave interferometer by
introducing an additional external potential into one of the two branches of the
splitted wave packet. Hence the band transitions act as beam splitters in a
Mach-Zehnder interferometer setup. These aspects will be discussed in more detail in a future
publication \cite{06manipulation}.

\section*{Appendix}
The time independent Schr\"odinger equation in the basis of Bloch waves consists of two coupled 
differential equations (\ref{G2_64}) and (\ref{G2_65}). Using the ansatz
\be
\langle\chi_{0,\kappa}\ket{\Psi}=Q_\kappa\,\rme^{-\rmi \kappa E/F } \qquad \mbox{and} \qquad
\langle\chi_{1,\kappa}\ket{\Psi}=P_\kappa\,\rme^{-\rmi \kappa E/F } \label{ansatz1}
\ee
we will show that in the case of a two-band system there exist exactly two energy ladders. 
A similar proof can be found in \cite{Fuku73} for a different system. 
Substituting (\ref{ansatz1}) into equations (\ref{G2_64}) and (\ref{G2_65}) we obtain
\be
\frac{\partial}{\partial \kappa}
\left(\begin{array}{*{2}{c}} 
Q_\kappa \\
P_\kappa \\ 
\end{array} \right)
=\frac{\rmi}{F}
\left(\begin{array}{*{2}{c}} 
E_{0,\kappa}-dF & M_\kappa^* \\
M_\kappa & E_{1,\kappa}\\ 
\end{array} \right)
\left(\begin{array}{*{2}{c}} 
Q_\kappa \\
P_\kappa \\ 
\end{array} \right)
=\mathbf{A}_\kappa 
\left(\begin{array}{*{2}{c}} 
Q_\kappa \\
P_\kappa \\ 
\end{array} \right) \,.
\label{G2_74}
\ee
The dispersion relations $E_{0,\kappa}$, $E_{1,\kappa}$ and the reduced transition matrix 
element $M_\kappa$ are $\pi/d$-periodic. Therefore we can apply Floquet's theorem, 
which says that a system of linear homogeneous differential equations with periodic coefficients 
as given above is solved by a fundamental system of the structure
\be
\mathbf{X}_\kappa=\mathbf{Y}_\kappa\,\rme^{\rmi\kappa\mathbf{Z}}\label{G2_75} \, ,
\ee
where $\mathbf{Y}_\kappa$ is a $\pi/d$-periodic $2\times2$\;-\,matrix and $\mathbf{Z}$ is a 
constant complex $2\times2$\;-\,matrix. Let the eigenvalues of $\mathbf{Z}$, 
called characteristic exponents, be denoted by $z_0$ and $z_1$.
They are only defined up to multiples of $2d$. Multiplication of the fundamental system 
(\ref{G2_75}) with the eigenvectors of $\mathbf{Z}$ from the right leads to
\be
\left(\begin{array}{*{2}{c}} 
Q_{\kappa,n} \\
P_{\kappa,n} \\ 
\end{array} \right)
=\rme^{\rmi \kappa z_n}
\left(\begin{array}{*{2}{c}} 
f_{\kappa,n} \\
g_{\kappa,n} \\ 
\end{array} \right)
\ee
with $\pi/d$-periodic functions $f_{\kappa,n}$ and $g_{\kappa,n}$. 
Here, $n=0,1$ distinguishes both possible eigenvectors. Because $\langle\chi_{0,\kappa}\ket{\Psi}$ and 
$\langle\chi_{1,\kappa}\ket{\Psi}$ have to be $\pi/d$-periodic, too, the phases are given by
\be
-\rmi \frac{\pi}{d} \frac{E}{F}+\rmi \frac{\pi}{d} z_n=-\rmi 2\pi m  
\ee
with $m\in\mathbb{Z}$. The energies are therefore 
\be
E_{n,m}=2mdF+Fz_n \,.
\label{G2_78}
\ee
Obviously, there exist exactly two different energy ladders as long as the eigenvalues 
of $\mathbf{Z}$ are distinct. Due to the fact that the Hamiltonian (\ref{S}) is hermitian, 
the energies $E_{n,m}$\,, and thus the $z_n$, are real numbers. 

\section*{Acknowledgements}
Support from the Deutsche Forschungsgemeinschaft
via the Graduiertenkolleg  ``Nichtlineare Optik und Ultrakurzzeitphysik''
and the Studienstiftung des deutschen Volkes
is gratefully acknowledged. 

\section*{References}

\bibliographystyle{unsrtot}

\end{document}